\documentclass[aps,prd,preprint,superscriptaddress,showkeys]{revtex4}

\usepackage{amsmath}
\usepackage{amssymb}
\usepackage{amsfonts}
\usepackage{upgreek}
\usepackage{color}
\usepackage{graphicx}
\usepackage{blindtext}
\usepackage{subfig}
\usepackage{amsthm}
\usepackage{physics}
\usepackage{bm}
\usepackage{slashed}
\usepackage{amsfonts}
\usepackage{booktabs}
\usepackage{siunitx}
\usepackage{listings}
\usepackage{xcolor}
\usepackage{booktabs}
\usepackage[utf8]{inputenc}
\inputencoding{latin1}
\inputencoding{utf8}

\newcommand{\sgn}{\text{sgn}}
\DeclareMathOperator{\arctanh}{arctanh}
\begin{document}
%\preprint{version 1.0}
\title[]{Kink solutions in a generalized scalar $\phi^4_G$ field model   }
\author{Jonathan Lozano-Mayo}
\email[Email:]{jonathanloz@ciencias.unam.mx}
\author{Manuel Torres-Labansat}
\email[Email:]{torres@fisica.unam.mx} 
\affiliation{Instituto de F\'{\i}sica,
Universidad Nacional Aut\'onoma de M\'exico,
Apartado Postal 20-364, Ciudad de M\'exico  01000, M\'exico}
\date{\today}
%\pacs{}
\begin{abstract}
We study  a scalar field   model  in a two dimensional space-time with a generalized $\phi^4_G$  potential which has four minima, obtaining novel kink  solutions with well defined properties although the potential is non-analytical at the origin. The model contains a control parameter   $\delta$ that breaks the degeneracy of the potential minima, giving  rise to two different phases for the system. The $\delta<0$ phases do not possess  solitary wave solutions. At the transition point $\delta=0$ all  the potential minima are degenerate and  three different kink solutions result. As the transition to the  $\delta>0$ phase takes place, the minima of the potential are no longer degenerate and a unique   kink  $\phi_\delta$ solution is produced.  Remarkably, this kink is a coherent structure that results from the merge of three kinks that can be identified with those observed at the transition point. To support the interpretation of $\phi_\delta$ as a bound state of three  kinks, we   calculate  the force between  the  kink-kink pair components of $\phi_\delta$, obtaining an expression that has  both  exponentially repulsive  and constant attractive contributions that  yields an equilibrium configuration,  explaining the formation of the  $\phi_\delta$ multi-kink state. We further investigate kink properties including their stability guaranteed by  the positive defined spectrum of small fluctuations around the kink configurations.  The findings of our work together with a semiclassical WKB quantization, including the one loop mass renormalization,  enable computing quantum corrections to the kink masses. The general results could be  relevant to the development of effective theories for non-equilibrium steady states and  for the understanding of the formation of coherent structures.
 \end{abstract}


\maketitle
%\tableofcontents
%%%%%%%%%%%%%%%%%%%%%%%%%%%%%%%%%%%%%%%%%%%%
\section{Introduction}\label{intro}

Solitons and  solitary waves are  remarkable properties of nonlinear field theories. 
Solitary waves are finite energy non-dispersive localised solutions of classical  field equations of motion   (\cite{Coleman1977,Rajaraman1982,Manton2004,Weinberg2012}).  When two solitary waves collide and each  preserves its form after scattering, we refer to them as  solitons  \cite{Zabusky1965}. Solitons and  solitary waves frequently display particle-like properties and are relevant to the understanding of a plethora of non linear  phenomena  in many areas of physics, for example: hydrodynamics \cite{Scott2003},  nonlinear optics \cite{Chen2012}, condensed matter \cite{Salomaa1987,Thouless:1998ww}, nuclear physics \cite{Skyrme1962},  quantum field theory \cite{Marciano1978}, and cosmology \cite{Zurek1996}.

Solitary waves satisfy the Euler-Lagrange  equations of  motion, yet it is also  essential to identify the stability condition  in many cases  related to the existence of conserved charges of topological origin  \cite{Finkelstein1968}.  In the case of two dimensions (space and time), the  scalar field theory with a $\phi^4$ potential gives rise to the kink solution \cite{Zeldovich1974,Vachaspati2006,Campbell2019}, that owes its stability to the existence of two degenerate minima in the potential; the solution approaches   different minima   as the field comes near to  spatial infinity in different directions. Also to be highlighted is the importance of  the $\phi^4$ model   for its connection to the study of phase transitions \cite{Landau1937}, the Ginzburg Landau theory of superconductivity \cite{Ginzburg1950,Tinkham1996}, and  the spontaneous symmetry breaking and Higgs mechanism  \cite{Kibble2015,Higgs1964,Englert1964}.

  The study of kinks  in $\phi^{2n}$ models with $ n \ge 3$ has recently attracted considerable attention \cite{ Gufan1978,Dorey2011,Gani2020} given that  the number and properties of kink solutions are extended by including   polynomial  potentials that have a  greater number of  minima.   These models can be used to analyse systems with multiple phase transitions, in which  it is possible to  describe successive first order  alternating with  second order   phase transitions   \cite{Khare2014}. Another finding is that in some cases the  kink-kink and kink-anti-kink forces  present  a polynomial  behavior  with respect to the separation  \cite{Manton_2019,Christov2019}, as compared to  the exponential  short range expression that is observed  in the $\phi^4$ model.

In this work we consider a  generalized $\phi^4_G$ model that possesses four inequivalent  minima, resulting from the addition of  non-analytical odd  powers of $ \abs {\phi}^n, \, n < 4 $ to the potential. In another context similar models  have been  used to study first order phase transitions \cite{Gufan2006,Fox1979},  and  it was recently found that a Landau theory for non-equilibrium  steady-states can be constructed if one exempts the assumption of analyticity in the effective potential \cite{Aron2020}. However in this work we are interested in the kink solutions of the relativistic $\phi^4_G$  model. We  prove that although the $\phi^4_G$ potential is non analytical at the origin $\phi =0$,
  a careful treatment of the  potential discontinuities enables  kink solutions with well defined properties. The kinks obtained in the different phases of the model are studied in detail, including the existence of a static multi-kink that results  from the bound state of three kinks. 
  This result  is possible  because the kink-kink interaction between the components of the multi-kink is given by  a  confining  potential.
  Additionally using  the  scheme based on the semiclassical functional quantization including the one loop mass renormalization \cite{Dashen1974,Goldstone1975,Evslin2019,Aguirre2020} we calculate  the quantum  mass corcections for the  kinks of the  $\phi^4_G$ model.

The paper is organized as follows. Section \ref{model} contains a general review of the formalism required to study kink solutions in scalar field theories in one space dimension. Section \ref{kinks} deal with the description of the  generalized 
 $  \phi_G^4$ model and a detailed study of the various kink solutions that are obtained in the degenerate and non-degenerate potential minima regions of the theory. The calculation of the force  between the components of the  multi-kink  is carried out in section \ref{kint}. Section  \ref{Qmass} contains the stability analysis of the classical configuration and also the explicit calculation of the quantum mass corrections of the kinks. The final considerations are presented in section  \ref{conclu}.

\section{Framework model}\label{model}

\subsection{Generalized  $\phi^4_G$  model and its particle excitations  around the vacuum}\label{Gphi4}

We consider a  scalar field in two dimensions (one space and one time) described by the Lagrangian density 

\begin{equation}
 \label{eq:Lagrangian}
 \mathcal{L}=\frac{1}{2} (\partial_\mu\phi)(\partial^\mu\phi)- U(\phi), 
\end{equation}
where $\phi(t,x)$ is a real scalar field, $\mu = 0,1$ and we set $\hbar = c=1$. The field equation of motion that follows from  this Lagrangian is given as

\begin{equation}
\label{eq:eqphiT}
   \frac{\partial^2\phi}{\partial t^2}-\frac{\partial^2\phi}{\partial x^2}=-\frac{\partial U}{\partial\phi}.
   \end{equation}
The energy functional corresponding to the Lagrangian  (\ref{eq:Lagrangian}) is given by the following expression 

\begin{equation}
\label{eq:EnerTot}
     E\left[\phi \right] =\int_{-\infty}^{\infty} dx  \left( \frac{1}{2}\left(\frac{d\phi}{dt}\right)^2+  \frac{1}{2}\left( \frac{d\phi}{dx}\right)^2+U(\phi) \right) .
\end{equation}
As far as the field potential is concerned we propose the following  
\begin{align}
\label{eq:Potencial}
 U(\phi) &= U_0(\phi) + \Delta U(\phi),   \nonumber \\ 
  U_0(\phi) &=\lambda  \left(  \abs{\phi}-v_1 \right)^2 \left( \abs{\phi}-v_2 \right)^2 , \qquad \qquad \Delta U(\phi) = \lambda \delta  \left( \abs{\phi}-v_2 \right)^2. 
\end{align}
This potential is a generalization of the usual $ \phi^4$
 potential $ U_{\phi^4} = \lambda  \left(  \abs{\phi}^2 -v \right)^2 $. In fact  $ U_{\phi^4}$ is recovered  from Eq. (\ref{eq:Potencial}) if we set $v_2=-v_1=v$ and $\delta=0$. The case to be studied includes  odd and even powers of $ \abs {\phi}^n, \, n \le 4 $, and the symmetry breaking  associated with  inequivalent  potential  minima will appear when $ v_1> 0 $ and $ v_2> 0 $.  Additionally, the incorporation of the $ \delta $ parameter  breaks the degeneracy of the potential minima. Usually the odd powers of $ \abs {\phi} $ are not incorporated in a scalar field potential because they  lead  to  non-analytical terms in the theory \cite{Landau1937,Tinkham1996,Campbell2019}. However it will be demonstrated that  a correct  treatment of the discontinuities leads to a model with well defined properties. 

 The potential $U(\phi)$ in Eq. (\ref{eq:Potencial})   has  minima 
  at   $\phi= \pm V_1 \, ,  \pm V_2 $; where $ V_1 \,, V_2 $  and the value  of the  maxima   $ V_m $ are given by the following expressions 

\begin{equation}
  \label{eq:minV}  V_2= v_2, \qquad  {V_m,V_1}=\frac{(3v_1+v_2)\pm\sqrt{(v_1-v_2)^2-8\delta}}{4},
\end{equation}
where  $ V_m (V_1) $  correspond to the  $+(-)$ selection. Throughout this work we consider $v_2 > v_1$, in such a way that the extrema points of the potential satisfy  $ V_1 \, < V_m <  \, V_2 $.  We assume that $U(\phi)$ has four different potential minima ($V_m$ and $V_1$ are real), hence the  $ \delta$ parameter must satisfy the conditions  

\begin{equation}
\label{eq:condiciondelta}
  - v_1 (v_1 + v_2)   < \delta<\frac{(v_1-v_2)^2}{8}.
\end{equation}
The potential in  Eq. (\ref{eq:Potencial}) gives rise to the spontaneous symmetry breaking of the discrete symmetry 
$\phi(x) \to - \phi(x)$, which can be verified considering  excitations around any of the minima defining $\phi(x) = V_i + \sigma_i(x)$. Plugging this expression into Eq. (\ref{eq:Lagrangian}) yields  two independent Lagrangians $\mathcal{L}_i$, the  details of which are quoted in the appendix (\ref{Ap1}).
As expected $\mathcal{L}_i$ do not contain linear terms in $\sigma_i(x)$, but  they include a quadratic mass term and also cubic and quartic interaction terms. The masses of the two normal modes are determined from the second derivative of the potential evaluated at  $\phi= \pm V_i$: 
\begin{equation}
\label{eq:m12}
 m_{1} =  \sqrt{4\lambda(V_1-V_m ) (V_1-V_2)},  \qquad m_{2}=\sqrt{4\lambda(V_2-V_1 ) (V_2-V_m )},
\end{equation}
where $m_1, \, m_2$ correspond to the masses of the  particle excitations around $\pm V_1$ and $ \pm V_2$ respectively.  When the cubic and quartic terms in $\mathcal{L}_i$ are neglected,  the dynamics of the system  is described by a  set of uncoupled harmonic oscillations, with  eigenvalues $ \omega^2_i =  m_i^2 + k^2 ,  \,i=1,2$ and plane wave solutions $\sigma_i = c_i \, \exp (i (\omega t -k x))$.  A perturbative incorporation of  the cubic and quartic terms in the formalism enables  the calculation of higher order effects.

\subsection{2-dimensional topological solitary waves}\label{2DTS}

We now focus on  time independent field configurations. The existence of stable solitary wave solutions requires that the potential have two or more degenerate minima.  Furthermore, in order that Eq. ($\ref{eq:EnerTot}$) be finite, its integrand   should vanish as $x$ goes to either plus or minus infinity. This implies that as $x \to \pm \infty$ the field must approach one of the minima $V_i$ of the potential and also that  $U(V_i)=0$, hence a stable solitary wave is obtained when the field  $\phi(x)$ interpolates between two different contiguous absolute  minima $V_i \ne V_j$.   By multiplying the time-independent version of   Eq. (\ref{eq:eqphiT}) by $d \phi /dx$ and  integrating, a first order Bogomolny  equation  \cite{Bogomolny1976} is  obtained 
 
 \begin{equation}
\label{eq:eqphi}
  \frac{1}{2}  \left(  \frac{\partial\phi}{\partial x}  \right)^2= U (\phi),
   \end{equation}
 where, according to the previous discussion, the integration constant is zero. 
Eq. (\ref{eq:eqphi}) describes a system that is mathematically identical to the problem of  a unit mass particle  with null total energy, that moves in a $-U (\phi)$ potential; the equivalent ``position'' and  ``time'' correspond to $\phi$ and $x$ respectively \cite{Coleman1977}. Eq. (\ref{eq:eqphi}) leads to   
 
 \begin{equation}
   \label{eq:relxphi} x - x_0 =\pm\int_{\phi_m}^{\phi(x)}  \frac{d\phi}{\sqrt{2U(\phi)}},
\end{equation}
  where the $+(-)$ signs correspond to the kink (anti-kink),  and the kink position  $x_0$  is arbitrary because of  the translational invariance symmetry of the Lagrangian. In order to get a consistent solution  $\phi(x_0)= \phi_m$ is selected as $\phi_m=V_m$, where $V_m$ is   the field value between $V_i$ and $V_j$ at which the potential acquires its maxima value.
  
The stability of the kink results from   the existence of a non-trivial topological charge that can be assigned to each configuration. In two dimensions the topological current is defined as $J_\mu= C   \, \epsilon_{\mu \nu}  \partial^ \nu  \phi(x)$, where $C$ is a constant and $ \epsilon_{\mu \nu} $ is the two dimensional Levi-Civita symbol. The current is automatically conserved because $ \epsilon_{\mu \nu} $  is antisymmetric $\partial^ \nu   \, J_\mu= 0$. The corresponding conserved charge is 

\begin{equation}
  \label{eq:Qtop}  Q= \int_{-\infty}^{\infty} J_0 \, dx =   \frac{1}{2 v_2}  \left(  \phi(+\infty) -   \phi(-\infty) \right),
\end{equation}  
where we selected $C = 1/ 2v_2$. 

\section{Kinks in the generalized  $\phi^4_G$  model  }\label{kinks}

As mentioned, kink solutions are obtained when  $\phi(x)$ interpolates between two  contiguous absolute  minima $V_i \ne V_j$ with $U(V_i) = U( V_j)=0$. The previous condition cannot be satisfied when  $ \delta < 0$.  Bubble solutions  may be obtained \cite{Barashenkov1988} in this region, but the  asymptotic conditions $\phi(+\infty)  =\phi(-\infty) = 0 $ yield a vanishing topological charge, implying that the bubbles are unstable. The cases of interest appear with the phase transition to positive  $\delta$ values. The  $\delta=0$ and $\delta>0$  regions  show markedly different properties and will be considered separately.

\subsection{Kink solutions:  degenerate minima case  ($\delta=0 $)  }\label{kABC}

We first look at the potential  $ U_0(\phi)$ in Eq. (\ref{eq:Potencial}),  obtained when $ \delta = 0 $. In this case the minima of the potential are located at $ \pm v_1 $ and $ \pm v_2 $ and they are all degenerate since $ U_0(\pm v_1) = U_0(\pm v_2) = $ 0. Therefore, we expect kink solutions that interpolate between the following pairs of potential minima: $ (- v_2, -v_1), (- v_1, v_1) $ and $ (v_1, v_2) $, in addition to the corresponding anti-kinks that invert the direction in which the potential minima are connected. 

Consider first the kink $ \phi_A $ localised in the topological sector $ (-v_2, -v_1) $. Substituting
 the potential $ U_0(\phi)$ into Eq. (\ref{eq:relxphi}) and taking into account that  $ \phi_m = -(v_1 + v_2)/2$, it is straightforward to integrate  
 Eq. (\ref{eq:relxphi});  inverting the result to obtain 
\begin{equation} 
 \label{eq:phiA}  \phi_A(x)=  - \frac{v_1 + v_2}{2}  + \frac{v_2 - v_1}{2}  \tanh \frac{m x}{2}.
\end{equation}
 Here  the central position of the kink was selected at $x=0$, and   the masses of the scalar excitations   in  Eq. (\ref{eq:m12}) reduce to $m= m_1= m_2= \sqrt{2 \lambda}(v_2-v_1)$.   According  to  Eq. (\ref{eq:relxphi})  the corresponding anti-kink configuration is obtained as  $\phi_{\bar A}(x) = \phi_A(-x)$.
The energy density ${\cal E}(x) =(d\phi/dx)^2$ is directly calculated using  Eq. (\ref{eq:phiA}) to obtain 
\begin{equation} 
 \label{eq:densEA} {\cal E}_A(x)=\frac{m^4}{ 32 \, \lambda  } \sech^4 \frac{m x}{2}.
\end{equation}
The kink mass $M_A$ is obtained substituting Eq. (\ref{eq:phiA}) into Eq. (\ref{eq:EnerTot}), whereas the topological charge is computed from  Eq. (\ref{eq:Qtop}),  resulting in 
\begin{equation} 
 \label{eq:MQA} M_A=\frac{m^3}{ 12 \, \lambda  } ,  \qquad  \qquad Q_{A,{\bar A}} =   \pm \frac{1}{2}  \left(1 - \frac{v_1}{v_2}  \right)  . 
\end{equation}
We point out that   $ \phi_A$ and    $\phi_{\bar A}$  have the same mass value, whereas  their charges have opposite signs. 

For the  kink $ \phi_C$  in the topological sector $ (v_1, v_2) $ the calculations are completely analogous. The kink profile is given by
$ \phi_C(x) =  \phi_A(x) + v_1 +  v_2$ and the corresponding mass and topological charge coincide with those of $ \phi_A$: 
$M_C=M_A$, $Q_{C,{\bar C}}=Q_{A,{\bar A}}$ .

\begin{figure}[h]
 \begin{center}
\includegraphics[width=12cm]{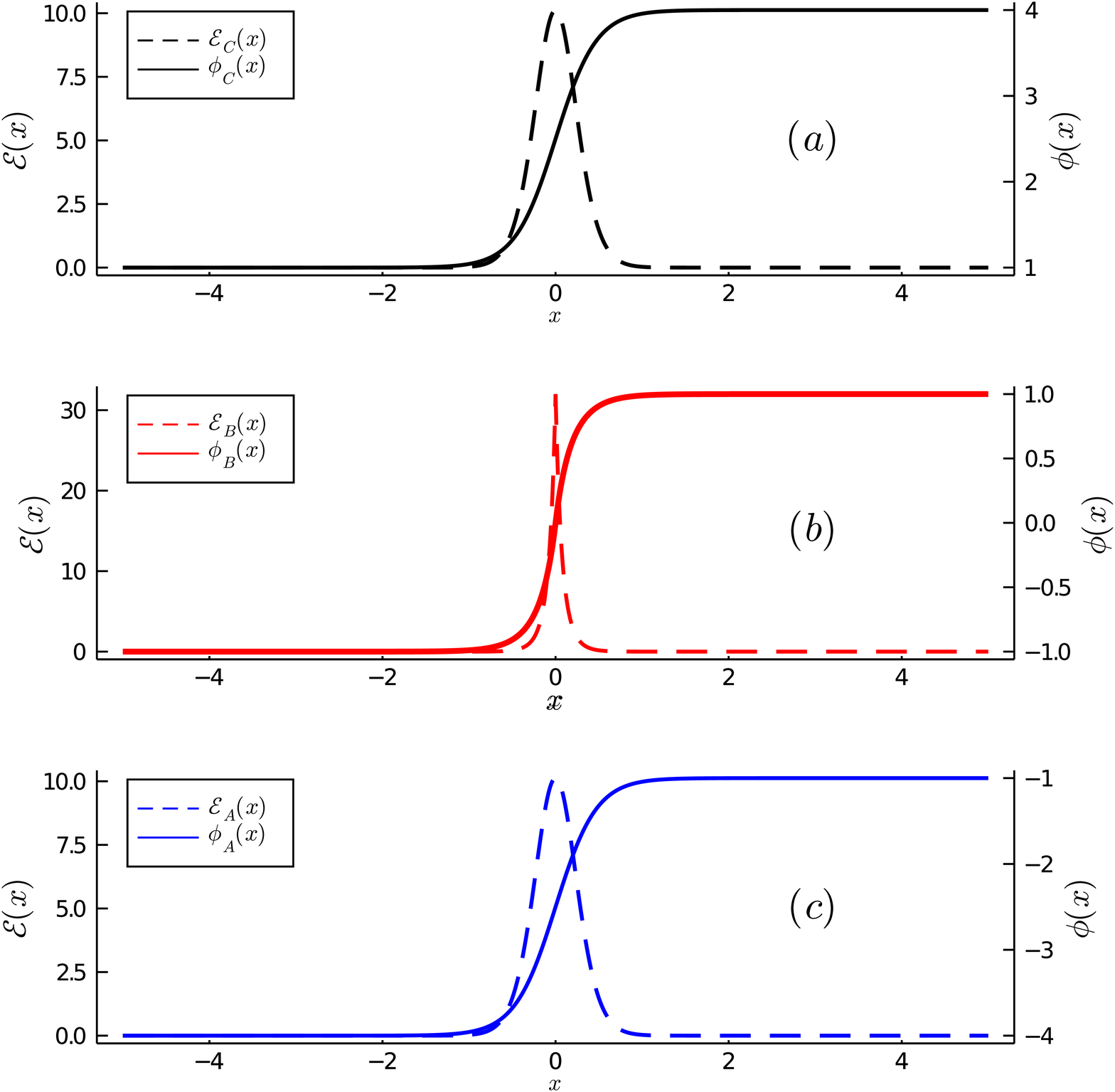}
\caption{Field solutions (continuous lines) and energy densities (dashed lines) for the kinks $ A$ (blue), $B$ (red) and $C$ (black). The parameters are selected as follows: $v_1=1 , \, v_2=4, \lambda =1, \delta=0.$
}
\label{KABC}
\end{center}
\end{figure}

The  kink  $\phi_B(x)$ is defined in the topological sector $ (-v_1, v_1) $  thus   changes sign, hence we must perform separate calculations for positive and negative values of $\phi_B(x)$. We  evaluate Eq. (\ref{eq:relxphi})  separately in the intervals $ (-v_1, 0) $   and  $ (0,v_1) $, considering that  $ \phi_m = 0$ and  selecting $x_0=0$. After evaluating the integrals, the results  can be inverted and written in  a single equation using a   $y$ variable  that is  piecewise defined as follows 
   \begin{equation}
\label{eq:defy}
 y =  \frac{m x}{2}
   + \frac{1}{2} sgn(x) \log \frac{v_2}{v_1} ,
      \end{equation}
where $ sgn(x)$ is the sign function. The result for   $\phi_B(x)$ is then given by 
\begin{equation}
   \label{eq:phiB} \phi_B(x)=    sgn(x) \frac{v_1 + v_2}{2}  + \frac{v_1 - v_2}{2}  \coth \left(  y \right) .
\end{equation}
Taking  into account that  $ \coth \left(   \frac{1}{2} \log \frac{v_2}{v_1} \right) = (v_1 + v_2)/(v_2 - v_1)$, both $\phi_B(0)=0$ and $\phi_B^\prime(0)= 2v_1 v_2/(v_2 - v_1)$ are continuos at the kink position; however  the second derivative is discontinuous.  We recall that $\phi_{\bar B}(x)= \phi_B(-x)$. utilising  Eq. (\ref{eq:phiB})  for $ \phi_B(x)$ the energy density is obtained
 \begin{equation} 
 \label{eq:densEB} {\cal E}_B(x)= \frac{m^4}{ 32 \, \lambda  } \csch^4 \left( y \right).
\end{equation}
The mass and topological charge are calculated as 
\begin{equation} 
 \label{eq:MQB} M_B=\frac{m^3}{ 6 \, \lambda}\frac{Q_B^2 (3- \abs{Q_B})}{ (1- \abs{Q_B})^3  } ,  \qquad  \qquad Q_{B,{\bar B}} =    \pm \frac{v_1}{v_2}. 
\end{equation}
 Fig. (\ref{KABC})   displays the plots for the three kink solutions and the corresponding energy densities. We observe that the profiles of the  three field configurations display a  characteristic kink behavior. The energy densities for $\phi_A(x)$ and $\phi_C(x)$ are smooth energy packets localised around the kink position, whereas the energy density for $\phi_B(x)$ presents  a spike maxima because  the second order derivative  of $\phi_B(x)$  is discontinuous at  the kink position.

We  point out  that we can define two parity symmetry operations $P_x$ and $P_\phi$ that invert the coordinate or field sign respectively: $P_x \phi(x) = \phi(-x)$ and  $P_\phi \phi(x) = - \phi(x)$. The first symmetry transforms any kink  into its corresponding anti-kink 
$P_x \phi_K(x) = \phi_{\bar K}(x) $. In the case of the $ \phi_B(x)$ both symmetries coincide.  However $P_\phi$  transforms the kink defined on  $(-v_2,-v_1)$ sector into  the anti-kink of the 
$(v_1,v_2)$ topological sector as follows 
\begin{equation}
 \label{eq:Pphi}
 P_\phi \, \phi_A(x) = -\phi_A(x) = \phi_{\bar C}(x) \, , \qquad \qquad P_\phi  \phi_{\bar A}(x) = \phi_C(x).
\end{equation}
We call $P_x$ and $P_\phi$  the charge and mirror symmetries.

\subsection{Kink solution for non degenerate  minima ($\delta \ne 0 $) }\label{kdelta}
   
\begin{figure}
 \begin{center}
\includegraphics[width=12cm]{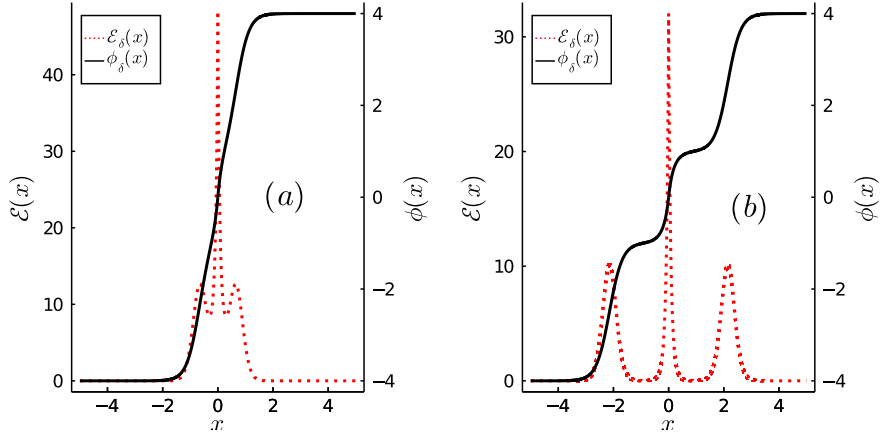}
\caption{Field  solutions $\phi_\delta(x)$ (continuous lines)   and their corresponding energy densities (dashed lines) for two values of the control parameter: $\delta=0.5$ (left plot) and $\delta= 0.001$ (right plot). The value of the other parameters are  $v_1=1 , \, v_2=4, \lambda =1$.
  }
\label{KD}
\end{center}
\end{figure}

 As the transition to the  $\delta >0$ phase takes place  the potential minima are no longer degenerate, $U(\pm V_1) > U(\pm V_2) =0$.  There are now only  two absolute minima,  hence a unique kink  interpolates between $\phi(-\infty) = - V_2$ to $\phi(\infty)  =  V_2$,  as well as its corresponding anti-kink. To determine the kink  configuration  we substitute the  potential  Eq. (\ref{eq:Potencial})   into Eq. (\ref{eq:relxphi}) and perform the integrations  in the intervals $ (0, v_2) $ and $ (- v_2,0) $ separately. In the entire interval $ (0, v_2) $ the field $ \phi $ is positive, considering that   $\phi_m=0$, and  selecting the kink position at $x_0=0$  we find 
 \begin{equation}
\label{eq:xphi}x =\int^{\phi}_{0}  \frac{d\phi}{\sqrt{2 U(\phi)}}  =       
\frac{1}{m_2}\log(\frac{ ( m_2 /  \sqrt{2} \lambda) \sqrt{(v_1-\phi)^2+\delta}+(\phi-v_1)\Delta_v+\delta}{(v_2-\phi)}) -g_0,
\end{equation}
 where   $\Delta_v= v_2-v_1$ and  $g_0$ is defined as 
\begin{equation}
    g_0=  \frac{1}{m_2}   \log(\frac{ ( m_2 /  \sqrt{2} \lambda) \sqrt{v_1^2+\delta}-v_1 \Delta_v +\delta}{ v_2}).
\end{equation}
An analogous expression is obtained for $x \le 0 $. Both relations can be explicitly inverted  and the results are summarized in the following expression 
\begin{equation}
 \label{eq:phid}
    \phi_{\delta}(x)= \sgn (x) \left(\frac{ v_2 \, e^{2 m_2(\abs{x}+g_0)}+ 2(v_2v_1-v_1^2-\delta) \, e^{m_2(\abs{x}+g_0)}-v_2\delta}{ e^{2 m_2(\abs{x}+g_0)}+ 2 \, \Delta_v \, e^{m_2(\abs{x}+g_0)}-\delta}\right).
\end{equation}
We can directly verify that  $ \phi_{\delta} (x) $ and its first derivative are continuous everywhere, but its second derivative is discontinuous at the origin ($ \phi (0) = 0 $) in agreement with Eq. (\ref{eq:eqphiT}) and the discontinuity that $\partial U / \partial \phi$ presents at that point. 
 The topological charge for  $ \phi_{\delta}$ is given as    $Q_\delta = \pm 1$.  The kink mass is evaluated   using Eqs. (\ref{eq:EnerTot},\ref{eq:eqphi})  with the following result
 \begin{align}
\label{eq:mdelta}    M_\delta& =\int_{-v_2}^{v_2}{\sqrt{2\lambda[(\abs{\phi}-v_1)^2+\delta](\abs{\phi}-v_2)^2}d\phi} \\ \nonumber &=\frac{\sqrt{2\lambda}}{3}\Bigg[\sqrt{\Delta_v^2+\delta}(\Delta_v^2-2\delta)- \,        \left( v_1(v_1-3v_2)-2\delta \right) \, \sqrt{v_1^2+\delta}\\ \nonumber &\qquad+ \, 3 \,  \delta  \Delta_v \log(\frac{\sqrt{\Delta_v^2+\delta}+\Delta_v}{\sqrt{v_1^2+\delta}-v_1})\Bigg].
\end{align}
Fig.(\ref{KD}) displays plots of  $  \phi_{\delta}(x)$ and the energy density $ {\cal E}_\delta(x)$ for two values of the $\delta$ parameter. For  small $\delta$,    $ \phi_\delta (x) $ clearly  shows  three successive kinks with values in the regions: $(-v_2,-v_1), \, (-v_1,v_1), (v_1,v_2) $. These kinks are centered at the positions $x=-x_m, \, x=0 , \, x=x_m$. The central position $(x=0)$ is arbitrarily selected because of the translational invariance symmetry of the system; however  $ x_m$ is fixed and its value will be explained below.  At  the positions of each of the internal kinks the energy density  shows a lump-like distribution, and in particular the central energy packet presents a spike  configuration. Based on these results we conclude that $ \phi_\delta $ is  a  bound state resulting from the merge of  the
 $\phi_A$, $\phi_B$ and $\phi_C$  kinks.  To   support this claim we  notice that the following charge equality is  trivially fulfilled
\begin{equation}
 \label{eq:Q=Q}
   Q_\delta \, = \, Q_A  \, + \, Q_B \, + \, Q_C = 1.
\end{equation}
Additionally,  it follows  that for small values of $\delta$  the kink mass in  Eq. (\ref{eq:mdelta}) can be approximated as 
\begin{equation}
 \label{eq:M=SM}
 M_\delta \approx M_A  \, + \, M_B \, + \, M_C + 
m \,  \delta \, \left( \frac{\vert Q_B \vert}{1-\vert Q_B \vert}   +\log \left[\frac{2 \, m^2 \, \vert Q_B\vert  }{ \lambda \, \delta ( 1- \vert Q_B \vert ) }    \right] \right) +{\cal O }(\delta^2)
\end{equation}
where we used the values for $M_A =M_C$ Eq. (\ref{eq:MQA}) and $M_B$ Eq. (\ref{eq:MQB}). The mass of the $\phi_\delta$ kink   results from the  addition of the  masses  of the constituent kinks and a term that, as shown in  section (\ref{kint}), represents the potential energy of the system evaluated at the equilibrium configuration Eq. (\ref{eq:M=SM}).

Using  Eq. (\ref{eq:xphi}) evaluated at $x=x_m$ and $\phi=V_m$  we obtain the  value of    $x_m$ that determines the distance  of $\phi_A$ and $\phi_C$ relative to $\phi_B$. For small  $ \delta $  it is approximated as
\begin{equation}
 \label{eq:xm}
 x_m  \approx  \frac{1}{ m } \, \log \left( \frac{ 2 \, \vert Q_1 \vert m^2 }{\lambda \, \delta  } \right)
 \,, \qquad \qquad   \delta  << \frac{m^2}{\lambda} .
\end{equation}
Hence $ \phi_\delta $ is a multi-kink state $ \phi_\delta(x) \approx 
  \phi_A(x+x_m) + \phi _B(x) + \phi_C(x-x_m) $, where according to  Eq. (\ref{eq:xm})  there is a large but finite separation.
  In section \ref{kint} we shall prove that  $\pm x_m$  represent the equilibrium positions of the forces  that act  on $\phi_A$  and $\phi_C$, considering that   $\phi_B$ is fixed at $x=0$.

  A fundamental property of solitary waves is that they are non-dispersive, namely they represent localised energy packages that move with constant speed, maintaining their initial structure.  To verify that the previously obtained solutions can be considered as true solitary waves,  we apply a boost by defining $ \phi_X (\gamma (x -v t)) $, with $ \gamma = 1/\sqrt{(1 - v^2)  } $  where $ \phi_X  $ is any of the kink solutions Eqs. (\ref{eq:phiA},  \ref{eq:phiB},\ref{eq:phid}).   We confim that   $ \phi_X (\xi) $  is a  solution of the  time dependent Eq. (\ref{eq:eqphiT}),  hence  the  four kinks ($ \phi_A,\, \phi_B,\,  \phi_C,\, \phi_\delta $) and their corresponding anti-kinks are indeed solitary waves.

\section{Kink interactions }\label{kint}

In this section we analyse the effect of combining two kinks or a  kink with an anti-kink, and compute the forces that act between them.  In  general these  configurations  will be  time dependent, but if we  consider  a $K K$ or $K { \bar K}$  pair that at  an initial time  is separated by a large distance  $R >> 1/m$, for a short period  it will experiment a rigid displacement and the force can be determined. In order to have a continuous  finite energy configuration it is required that a kink $K$ is  followed by an antinkink $ { \bar K}$ defined in the same topological sector or a  kink $K^\prime$ defined in a contiguous topological sector. Thus we consider three independent configurations: $C { \bar C}$,  $B { \bar B}$ and $ B C$ that will give rise to different expressions for the corresponding forces. These results will be extended to analyse the forces that act within the multi-kink $\phi_\delta$,

The  momentum density for a scalar field obtained from Noether's theorem is given as $ -\left(   \frac{ \partial  \phi}{  \partial  t } \right) \left(  \frac{ \partial   \phi}{  \partial x } \right)$. Integrating this expression in the interval $(X, \infty)$ and  utilising  Eq. (\ref{eq:eqphi})  the  force acting  on the field to the right of  $X$ is obtained as \cite{Manton2004,Manton_2019} 
\begin{equation}
\label{eq:F}
	F = \frac{d P}{dt}  = -  \frac{d }{dt} \int_X^\infty \left(   \frac{ \partial  \phi}{  \partial  t } \right) \left(  \frac{ \partial   \phi}{  \partial x } \right) dx = \left[ \frac{1}{2} \left( \frac{d \phi}{dx}  \right)^2 - U(\phi) \right]_{x=X}.
\end{equation}
Here we took into account that the  last two terms in the previous equation 
cancel out  as $x  \to  \infty$.

\subsection{$\delta=0$ case }\label{FABC}
 
\begin{figure}
 \begin{center}
\includegraphics[width=12cm]{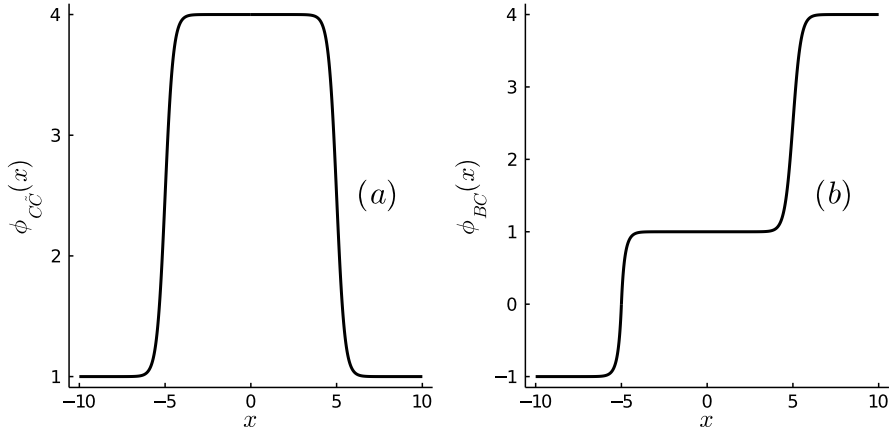}
\caption{(a) Multi-kink ansatz  $ \phi_{C { \bar C}} (x)=  \phi_C  (x+q)  +   \phi_{ \bar C}  (x-q) - v_2 $  used to calculate the  $C { \bar C}$ interaction.    (b) Multi-kink ansatz $\phi_{B { C}} (x)  =  \phi_B  (x+q)  +   \phi_{ C}  (x-q) - v_1 $ utilised to calculate the $BC$ interaction. The   parameters are selected as   $v_1=1 , \, v_2=4, \lambda =1, \delta=0\, , $ and  $q=5$. The kink separation is $R=2q=10$ in both cases.
  }
\label{MKansatz} 
\end{center}
\end{figure}

 First we analyse the $C { \bar C}$ configuration in which the kink $C$  localised  at $x= -q$ occupies the  $x < 0$ region, whereas  $ { \bar C}$ defined for  $x>0$ is situated at $x= q$; the ansatz for  the $C { \bar C}$ field is represented as $ \phi_{C { \bar C}} (x)=  \phi_C  (x+q)  +   \phi_{ \bar C}  (x-q) - v_2 $. When $q \gg 1/m$ we can approximate 
$\phi_{ \bar C}  (x-q) - v_2  \approx 0$  for $x<0$, and $\phi_C  (x+q) -v_2  \approx 0$ on the positive $x$-axis; hence  $\phi_{C { \bar C}} $ is a correct representation for the $C{ \bar C}$ system as shown in  Fig.(\ref{MKansatz}). Thus, if we select $X=0$  in Eq. (\ref{eq:F}) $F$  represents the force exerted on ${\bar C}$.    Selecting $x <<q $ and $q \gg 1/m$ we obtain   $\phi_{C { \bar C}} (x)  \approx   v_2 + 
2 (v_1-v_2) \, e^{-m q }+ {\cal O }(x^2)$, plugging this expression  into Eq. (\ref{eq:F}) yields
  $\left( \frac{d \phi_{A { \bar A}}  }{dx}  \right)^2_{x=0}= 0 $ and  $U_0(\phi_{A { \bar A}})_{x=0} = \frac{m^4}{\lambda}   e^{-2mq}  $, that results in an attractive interaction with an exponential decay given as follows

\begin{equation}
\label{eq:FCbC}
	F_{C { \bar C}} \approx     -  \frac{m^4}{\lambda} \, e^{- m\, R}   , 
\end{equation}
where $R= 2 q$ is the separation between the kink and the anti-kink.

In order to compute $F_{B { \bar B}}$  we define $\phi_{B { \bar B}} (x)=\phi_B  (x+q)  +   \phi_{ \bar B}  (x-q) - v_1$. Using $\phi_B(x)$ given in Eq. (\ref{eq:phiB}),  the following asymptotic expression $(x <<1 \, , \, q \gg 1/m)$ follows up  $\phi_{B { \bar B}} (x)  \approx   -v_1 + 2 \frac{v_1}{v_2}(v_2-v_1) \, e^{-m q }+ {\cal O }(x^2)$,   from which we obtain
$\left( \frac{d \phi_{B { \bar B}}  }{dx}  \right)^2_{x=0}= 0 $ and  $U_0(\phi_{B { \bar B}})_{x=0} = (\frac{v_1}{v_2})^2 \frac{m^4}{\lambda}   e^{-mR}  $, that also produce an attractive interaction but with different strength

 \begin{equation}
\label{eq:FBbB}
	F_{B { \bar B}} =    -  Q_B^2  \, \frac{m^4}{\lambda} \,  e^{- m\, R}   . 
\end{equation}  

We now turn the attention to the  $BC$ system. The field ansatz configuration is written  as 
$\phi_{B { C}} (x)  =  \phi_B  (x+q)  +   \phi_{ C}  (x-q) - v_1 $, Fig.(\ref{MKansatz}) shows that $\phi_{B { C}} (x)$ represents a field that interpolates between the $-v_1$ and  the $+v_2$ potential minima.  When $x <<1 \, , \, q \gg 1/m$ the asymptotic expression for $\phi_{B { C}} (x)$ reduces to 

\begin{equation}
\label{eq:phiBC}
	\phi_{B { C}} (x) \approx   -v_1 + \frac{(v_2-v_1)^2}{v_2}\, e^{-m q }+ 
	m  \, x \, \frac{(v_2^2-v_1^2)}{v_2}\, e^{-m q }   + {\cal O }(x^2).
\end{equation}
In this case we find 
$\frac{1}{2} \left( \frac{d \phi_{Bc}  }{dx}  \right)^2_{x=0}= (1+\frac{v_1}{v_2})^2 \frac{m^4}{4 \lambda}   e^{-mR} $ and  $U_0(\phi_{B C})_{x=0} =  (1-\frac{v_1}{v_2})^2 \frac{m^4}{4 \lambda}   e^{-mR}  $; that lead to a  repulsive interaction 

\begin{equation}
\label{eq:FBC}
	F_{BC} =   \abs{Q_B}  \frac{m^4}{\lambda} \, e^{- m\, R}.
\end{equation}  
The previous results show that generically  the kink-anti-kink  interaction  is attractive,  while the  kink-kink  interaction  is repulsive, and that in  both cases the interaction decays  exponentially as  the kink separation increases.  However the interaction strengths   are not equal in absolute value,  but rather the ratio of the forces are given as follows 

\begin{equation}
\label{eq:RF}
	\abs{ \frac{F_{B { \bar B}}}{F_{B C} }
 } =  \abs{Q_B}   ,  \qquad \qquad   \abs{ \frac{F_{C { \bar C}}}{F_{B C} }
 } =    \frac{ 1}{\abs{Q_B}   }. 
\end{equation} 
The forces for other  multi-kink configurations can be obtained from the relations    $F_{A { \bar A}} = F_{C { \bar C}}$, $F_{A B} = F_{B C}$.

\subsection{$\delta  > 0$ case }\label{FDelta}

When $\delta >0$,   $ \phi_A $, $ \phi_B $ and $ \phi_C $ are no longer exact solutions of Eq. (\ref{eq:eqphi}),  instead we have a new kink $ \phi_\delta $ that interpolates between the  two absolute potential minima $ \pm v_2 $. However, as mentioned in  section (\ref{kinks}),   the $ \phi_\delta(x)  $ profile resembles a multi-kink formed from the merge  of $ \phi_A $, $ \phi_B $ and $ \phi_C $, which leads us to analyse the configuration defined by the following ansatz
\begin{equation}
\label{eq:phiS}
	\phi_{ABC} (x;q)= 	 \phi_A  (x+q)  +   \phi_B  (x)+ \phi_C  (x-q).
\end{equation}
We compute the pair interactions between the  components of  $ \phi_{ABC} (x;q) $ and determine if it is possible to find a value of $ q $ for which the configuration is stable. 
The force exerted on  the field $\phi_C  (x-q)$ in Eq. (\ref{eq:phiS}) is dominated from the $\phi_B -\phi_C$ interaction, that was already calculated in Eq. (\ref{eq:FBC}). However that  calculation  was carried out for $\delta =0$, which  only includes   the  potential $ U_0(\phi)$ in Eq. (\ref{eq:Potencial}). Hence we must add the contribution from $ \Delta U (\phi)= \lambda \delta  \left( \abs{\phi}-v_2 \right)^2 $ evaluated at $ \phi_ {BC} (0) $ using Eq. (\ref{eq:phiBC}). A direct calculation yields  $ \Delta U (\phi_ {BC} (0))  \approx m^2 \delta /2$. Adding this result to  $F_{BC}$ given in
Eq. (\ref{eq:FBC})   the total force acting on $ \phi_ C $  results in 
 
 \begin{equation}
\label{eq:FABC}
	F_{AB;C} =    \abs{Q_B}  \frac{m^4}{\lambda} \, e^{- m\, q} -  \frac{m^2}{2} \delta. 
\end{equation}  
  Remarkably,  in addition to the repulsive contribution, there is an attractive constant long range force. Thus,  the dynamics of the  position of the kink $ \phi_C $ takes place in an effective potential  $V_{AB;C}(q) $ that is obtained from the space integral of Eq. (\ref{eq:FABC}) leading to 
 
  \begin{equation}
\label{eq:VABC}
	V_{AB;C}(q)  =  \abs{Q_B}  \frac{m^3}{\lambda} \, e^{- m\, q} +  \frac{m^2}{2} \delta  \, +  \frac{1}{2} K .
\end{equation}  
where $K$ is a constant. This potential has  a stable equilibrium point  $q = x_m$, that  coincides with the  separation between  the contiguous kinks components of  $ \phi_\delta $ given in Eq. (\ref{eq:xm}). As the force acting  on $ \phi_ A $ is identical to Eq. (\ref{eq:FABC}) and the the force on $ \phi_ B $ cancels, it follows that the effective potential for  the  
$\phi_{ABC}$ multikink is $V_{ABC}(q) = 2 V_{AB;C}(q) $. When this potential is evaluated at the equilibrium point  $q = x_m$ and selecting the constant in Eq. (\ref{eq:VABC}) as $K= m \delta \left(\frac{2 \vert Q_B \vert -1}{1 - \vert Q_B \vert} - \log (1 - \vert Q_B \vert) \right)$  it follows that $V_{ABC}(x_m)$ exactly coincides with the correction of order $\delta$ to the  $\phi_\delta$ mass   in Eq. (\ref{eq:M=SM}), confirming that it represents the potential energy of the multi-kink at the equilibrium configuration.

\begin{figure}
 \begin{center}
\includegraphics[width=12cm]{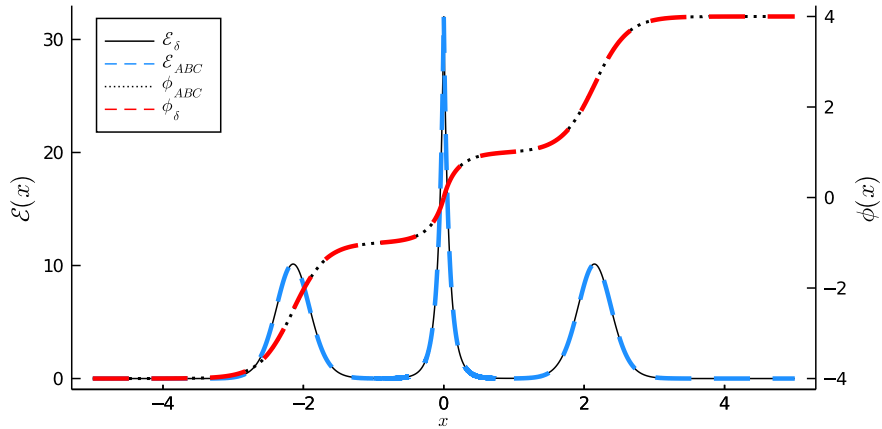}
\caption{
Comparison of the exact field configuration $\phi_\delta (x) $ Eq. (\ref{eq:phid})  dotted (black) line, with the multi-kink $\phi_A (x+x_m) + \phi_B (x)+ \phi_C (x-x_m)$ Eq. (\ref{eq:phid2}) dashed (red) line, where  $x_m$ is given in Eq. (\ref{eq:xm}). The corresponding energy densities are also compared.
As observed the  two configurations coincide for  small  $\delta$.  The parameter values are $ \delta = 0.001, \, v_1=1 , \, v_2=4, \lambda =1$. 
 }
\label{MKABC}
\end{center}
\end{figure}

 Finally, to complete the interpretation of $ \phi_\delta $ as a $ \phi_A \, +  \phi_B \, + \phi_C $ multi-kink state,  
 we notice that for $ \delta << 1$, both  $ \phi_\delta (x)  $  Eq. (\ref{eq:phid}) and  $ \phi_{ABC} (x;x_m) $ Eq. (\ref{eq:phiS})  reduce to  the same expression 
   \begin{equation}
 \label{eq:phid2}
    \phi_\delta (x)= \sgn (x) \left(\frac{ 4 \, v_2 \, \Delta_v^2 \, e^{2 m(\sgn (x)-x_m)} \, + \, 4 \, v_1 \,  \Delta_v^2 \, e^{m (\sgn (x)-x_m)} \, - \, v_2 \, \delta}{ 4 \, \Delta_v^2 \, e^{2 m(\sgn (x)-x_m)} \,+ \, 4   \, \Delta_v^2  \, e^{m (\sgn (x)-x_m)}-\delta}\right),
\end{equation}
 where  $\Delta_v = v_2 -v_1$.  Fig.(\ref{MKABC}) compares the plots obtained from the previous expression with the exact solution given in Eq. (\ref{eq:phid}). It is clearly shown that the difference between the two expressions becomes imperceptible as the  $\delta$ parameter  decreases.

%%%%%%%%%%%%%%%%%%%%%%%%%%%%%%%%%%%
 
 \section{Quantum mass corrections, and  renormalization }\label{Qmass}

 We now turn our attention to the study of field excitations around the kink configurations. The  kinks are  expected to be stable, this is verified by showing that the spectrum of the fluctuations is positive defined. Furthermore the use of a WKB approximation  enables  computing the quantum corrections to the kink mass. In what follows we  mainly focus on the case of degenerate potential minima $ \delta =0$, hence the field potential is given by $U_0(\phi)$ in Eq. (\ref{eq:Potencial}).
 
 The field  $\phi(x,t)$ is written as the sum of the classical kink solution 
 $\phi_K(x)$ and a small  fluctuation $\eta(x,t)$ as follows $\phi(x,t) = \phi_K(x) +  \eta(x,t)$. If we substitute the previous decomposition  into the energy functional Eq. (\ref{eq:EnerTot}), integrating by parts and taking into account that $\phi_K(x) $ satisfy Eq. (\ref{eq:eqphi}), we find that
 
  \begin{equation}
\label{eq:EnerTot2}
     E_K = M_K \,+  \int_{-\infty}^{\infty} dx \left\{ \frac{1}{2}   \left( \frac{d \eta }{dt }  \right)^2 +\eta \left(  -  \frac{d^2 \eta}{dx^2} + U_0^{ \prime  \prime}   ( \phi_K)  \eta + \delta(x-x_c) \, disc \,   \partial_x  \eta \,  \right)  + {\cal O} (\eta^3 ) \right\}. 
    \end{equation}
Compared with the usual results  \cite{Rajaraman1982,Dashen1974,Goldstone1975,Boya1989}, this expression includes an extra term that takes into account that  $ \partial \eta / \partial x$ is discontinuous at the point $x_c$ where $ \phi_K(x_c) = 0$. In Eq. (\ref{eq:EnerTot2}) $disc \,   \partial_x \eta$ is given by 

  \begin{equation}
\label{eq:disceta}
    disc \,   \partial_x  \eta    = \lim_{ \epsilon \to 0 }   \left\{ \left( \frac{\partial \eta }{ \partial x }\right)_{x_c + \epsilon} -  \left( \frac{\partial \eta }{ \partial x }\right)_{x_c  - \epsilon} \right\}.
    \end{equation}
 Taking the usual expansion $ \eta(x,t) =  \sum_{m} \eta_m(x) \, e^{i \omega_m t }  $, the energy in Eq. (\ref{eq:EnerTot2}) is  diagonalised if the functions $\eta_m(x)$ are selected as  eigenfunctions of the following one dimensional Schr\"{o}dinger equation 
 
\begin{equation}
\label{eq:eqsch}
     \left(  -  \frac{d^2 }{dx^2} + U^{ \prime  \prime}( \phi_K)  + D_m \, \delta(x-x_c)  \right) \, \eta_m (x)  =   \omega_m^2 \, \eta_m(x), 
    \end{equation}
where $D_m =  disc \,  \partial_x  \eta_m /   \eta_m(x_c)$. 
The fact  that  $U_0(\phi)$ in (\ref{eq:Potencial}) is a quadratic form permits making contact with the supersymmetric quantum mechanics formalism \cite{Boya1989,Cooper1995}. Writing  $U_0(\phi) = \frac{1}{2} V(\phi)^2  $ with $ V(\phi) = \sqrt{2 \lambda }  \left(  \abs{\phi}-v_1 \right) \left( \abs{\phi}-v_2 \right) $ results in  the superpotential  $W(\phi) $  and the SUSY-QM potentials $V^{\pm}_{S}  (x)$ being  expressed as follows
\begin{equation}
  \label{eq:SusyPot}   W(x) = \frac{\partial V}{\partial \phi }[\phi_K(x)], \qquad  \qquad V^{\pm}_{S}  (x) =  W^2(x)  \mp \frac{dW(x)}{dx}.
  \end{equation}
The potential   in the stability Eq. (\ref{eq:eqsch})  coincides with one of the SUSY-QM potentials $U_0^{\prime\prime}(\phi_K)  = V^{+}_{S}  (x)$. 

 The solutions  of Eq. (\ref{eq:eqsch})  apply both  (i)  when the  fluctuations take place around the vacuum configurations  $ \phi_V =  \pm V_{1,2}$, in which case 
$U^{ \prime  \prime} (V_{1,2}) )= m$  and  the solution is given by  $ \omega_{{k^\prime }}^2 =m^2 + k^{\prime \,2 } $ and $ \eta_{k^\prime } (x) = e^{i {k^\prime }x}$; and  (ii)  when    the background is the kink  configuration, the explicit solutions of which will be presented in following paragraphs, but generically  include a discrete $ \omega_n$ and  a continuous  
 $ \omega_k$  spectrum whose eigenfunctions   have  an asymptotic behavior $ \eta_k (x) = e^{i kx  \pm \delta(k)/2}$ as $x \to \pm \infty$,    where  $\delta(k)$ is the  phase shift. Notice that the  asymptotic behavior of  $ \eta_k (x)$ implies a  reflectionless potential. 
 
The contribution of the continuous modes  is computed incorporating a regularization scheme in which the system is enclosed in a finite box of length $L$. Hence  the kink  excitation spectrum $ \omega_j^2 = m^2  + k_j^2$,  and  that of the vacuum fluctuations  $ \omega_j^{\prime \, 2 }= + m^2 + k_j^{\prime \, 2 }$ becomes discrete.  Taking into account the periodic boundary conditions,  $ L k_j  + \delta(k_j) = L  k_j^\prime = 2  \,  \pi \, j$, we  calculate the zero point energy contribution to the kink mass, subtracting the vacuum energy, to obtain 

\begin{equation}
\label{eq:Mcont}
M_{cont} =  \frac{1}{2}   \sum_j  \left( \sqrt{  m^2  + k_j^2) }  - \sqrt{ m^2 +  k_j^{\prime \, 2 } } \right) \sim -  \frac{1}{2}  \sum_j \frac{1}{L}   \delta(k_j) \frac{\Delta \omega_j}{ \Delta k_j}
\sim -   \int_{- \infty}^\infty   \frac{dk}{4 \pi}    \frac{d \omega_k}{ d k} \delta(k) ,
\end{equation}
where in the last equality we set $  \frac{1}{L}  \sum_j  \to \frac{dk}{2 \pi}$ as ${L \to \infty}$. This quantity is still logarithmically divergent and it requires  the addition of a mass counter-term  obtained by a  one loop perturbative  renormalization scheme  \cite{Dashen1974,Aguirre2020}. The mass  counter-term takes the  form   $- \, \frac{1}{2} \, \delta  \,  m^2  \,  \langle { \cal V } \rangle$, where  the  tadpole contribution  and the average stability potential are given as

\begin{align}
 \label{eq:tadp}
  \delta m^2  &=  \,   \frac{6 \lambda }{ \pi} \, \int_{0}^\Lambda \,  \frac{dk }{  \left( k^2 + m^2 \right)^{1/ 2 } }  \, =  \,   \frac{6 \lambda}{  \pi}  \, \log  \frac{2 \Lambda}{  m} , \nonumber
\\      \langle {\cal V}  \rangle & =  \frac{1  }{ 12 \lambda}  \, \int_{-\infty}^\infty \, dx \, {\cal V}(x) ,  \qquad {\cal V}(x) = U^{ \prime  \prime}   ( \phi_K(x) ) -m^2  .
\end{align}
Here  $\Lambda$ is the momentum cut-off. Gathering  the contributions of the classical kink mass $M_K$, the discrete as well as the continuous zero energy modes, including  the subtraction of  the vacuum energy Eq. (\ref{eq:Mcont}), together with the kink renormalization term Eq. (\ref{eq:tadp}), we finally arrive to  the  following expression  for the quantum kink mass

  \begin{equation}
\label{eq:MKren}
    \tilde M_K= M_K \,+       \sum_n \,  \frac{1}{2} \, \omega_n - \int_0^\Lambda \, \frac{d k }{ 2 \pi } \, \delta(k)  \frac{d \omega_k }{ d k  } -  \, \frac{1}{2}   \,  \delta  \,  m^2  \,  \langle { \cal V } \rangle. 
      \end{equation}
  In the  next subsections we demonstrate that the formalism gives finite results for the  quantum masses of the 
  $\phi_A$, $\phi_B$ and $\phi_C$  kinks. 

%%%%%%%%%%%%%CCCCC%%%%%%%%%%%%%%%%%%%%%%
\subsection{$\phi_C$ quantum mass corrections  }\label{QmassC}

 In this subsection we compute the quantum mass  $ \tilde M_C$ for the  kink $\phi_C$.
Using the expression for   $ \phi_C(x)$  to evaluate the effective potential in Eq. (\ref{eq:eqsch}),  and taking into account that  $ \phi_C(x)$ is positive in the complete   $x \in ( -\infty,\infty)$ interval,  it follows that equation Eq. (\ref{eq:eqsch}) becomes  
\begin{align}
 \label{eq:eqschC}
 \left( -  \, \frac{1}{2} \, \frac{d^2 \eta_k}{dz^2} +  {\cal V}_C (z)  \right) \eta_k \,  &=   \epsilon_k \, \eta_k \,,
 \nonumber
\\       {\cal V}_C (z)  =  U_0^{ \prime  \prime}   ( \phi_C(z) ) -m^2  \, &=   \, -3 \, m^2 \,  \sech^2 \left( z \right)\, ,
\end{align}
 where  $z = mx /2$ and $\epsilon_k = 2 (\omega_k^2 -m^2)/m^2$.  Eq. (\ref{eq:eqschC})  is one of the SUSY  partner  equations  in   (\ref{eq:SusyPot})  with the superpotential  
 $W(z) = m \tanh (z)$.  The solutions to  Eq. (\ref{eq:eqschC})  are  well known \cite{Morse1953}. The spectrum consists of two discrete levels with energies and eigenfunctions given by:  $\omega^2_0 = 0$, $\eta_0(x)  =    \sech^2 \left( z \right)$; and $ \omega^2_1 =   \frac{3}{4} m^2$, $\eta_1(x)  =  \tanh \left(z \right) 
 \,  \sech \left( z\right)$. The continuous spectrum $  \omega^2_k =   k^2 +  m^2 $  has  the eigenfunctions $\eta_k (x)  = \, e^{i k x}\, H(\tanh z )$    where  the function $H(\xi)$ is defined as 
\begin{equation}
\label{eq:defHy}
H(\xi) = \left(
m^2 ( 3 \xi^2  -1 ) - 4 k^2 - \, i \, 6  \, k \, m \,  \xi \right) .
      \end{equation}
      The asymptotic behavior  of the continuous wave functions takes the following form $ \eta_k (x) = e^{i kx  \pm \delta_C(k)/2}$,   where  $\delta_C(k) = -2 \arctan 3 k m / ( m^2 - k^2)$. Plugging $\delta_C(k)$  into Eq. (\ref{eq:Mcont}), integrating by parts, and separating  the finite contribution from the one that diverge with $\Lambda$, leads  to 

\begin{equation}
\label{eq:McontC}
M_{cont-C} =  -  \frac{3m }{2 \pi  } -  \frac{6 m }{2 \pi  }  \int_0^\Lambda  \, dk \,   \frac{   (m^2 + 2 k^2 )}{  (m^2 + 4 k^2 ) \, \sqrt{m^2 +  k^2 }}  \, = - \frac{ m }{2 \pi  }  \left(  3 + \frac{ \pi }{ \sqrt{3}   } + 3  \log  \frac{2 \Lambda}{  m}\right).
\end{equation}
This quantity is still logarithmically divergent and we have to add the  mass counter-term indicated in Eqs. (\ref{eq:tadp},\ref{eq:MKren}). Taking into  account that  $ \langle {\cal V_C}  \rangle = - m/2  \lambda$,  the mass counter-term reduces to 
\begin{equation}
\label{eq:MrenC}
-   \, \frac{1}{2} \, \delta  \,  m^2  \,  \langle { \cal V_C} \rangle  \, =   \frac{3 }{2   \pi}  \, m \,   \log  \frac{2 \Lambda}{  m}  , 
\end{equation}
which exactly cancels the divergent term in Eq. (\ref{eq:McontC}).
 It is   now straightforward to add  the   contributions  in  Eqs. (\ref{eq:McontC},\ref{eq:MrenC})  with  the  classical kink mass and  the discrete energy  $ \frac{1}{2}  \omega_1$, to obtain the final result for the quantum $\phi_C$ kink mass 

   \begin{equation}
\label{eq:mCren}
    {\tilde M}_C=   \frac{m^3}{ 12 \lambda } + m \, \left( \frac{1}{ 4 \sqrt{3}  }  -\frac{3}{ 2 \pi }  \right) \, + \,  { \cal O }  (\lambda).
      \end{equation}
    The  same result is obtained for the  $A$ kink ${\tilde M}_A = {\tilde M}_C $.

    It is noteworthy  that  the stability equation (\ref{eq:eqschC}), its solutions, and the  quantum mass 
     (\ref{eq:mCren})  for the  $\phi_C$ kink,    coincide with the results previously obtained for the  $\phi^4$ model \cite{Dashen1974}, notwithstanding  the $ \phi^4 $ and $ \phi^4_G $ models are notoriously different. This can be  explained by  the reconstruction method \cite{Jackiw1977,Vachaspati2004},  in which the  structure of  the scalar field theory is obtained from  the  stability equations and the knowledge of the bound spectrum. In particular it has been show that when the spectrum has  two bound states, the reconstruction is not unique \cite{Bazeia2017}. Hence we conclude that considering the stability equation and its corresponding spectrum, the application of the reconstruction method  should  produce both the  $ \phi^4 $ and $ \phi^4_G $ models. 
 %%%%%%%%%%%%%%%%%%%%%%%%%%%%%%%%%%%%%

  \subsection{$\phi_B$ quantum mass corrections  }\label{QmassB}

  Consider now the  quantum fluctuations around the $ \phi_B$ kink. In Eq. (\ref{eq:phiB}) we recall that   $ \phi_B(x)$  is separately defined in the positive and negative $x$-axis.  Using the expression for   $ \phi_B(x)$  to evaluate the effective potential in Eq. (\ref{eq:eqsch}) leads to the  following   Schr\"{o}dinger equation 

\begin{equation}
\label{eq:eqschB}
        \left(  -  \frac{1}{2} \, \frac{d^2 }{dy^2} +   \, 3  \,   \csch^2 \left( y \right) + \frac{D_k}{m^2} \, \delta(x-x_c)   \right) \eta_k (y)   =   \epsilon_k \, \eta_k(y),
    \end{equation}
where  $y =  \frac{m x}{2}
   + \frac{1}{2} sgn(x) \log \frac{v_2}{v_1}$  was  piecewise defined in  Eq. (\ref{eq:defy}),  $\epsilon_k = 2 (\omega_k^2 -m^2)/m^2$, and the effective potential Eq. (\ref{eq:tadp}) is  now given as 
 $ {\cal V}_B (z)  =  U_0^{ \prime  \prime}   ( \phi_B(z) ) -m^2  \, =    \, 3  \, m^2  \csch^2 \left( y \right)$. Note that according to  Eqs. (\ref{eq:disceta},\ref{eq:eqsch})  the delta term potential  has to be included because $ \phi_B(x)$ cancels at $x=0$, so the derivative of $\eta_k$ is expected to be discontinuous at the origin.  Eq. (\ref{eq:eqschB}) is one of the SUSY-QM equation corresponding to the superpotential 
 $W(y) = m \coth y$.  
  
  We present the details of the solutions to Eq. (\ref{eq:eqschB}) in the appendix \ref{Ap2}. The spectrum and its corresponding eigenfunctions are given as 
    \begin{align}
\label{eq:specB} \nonumber  
     \omega^2_0 &= 0 \,, \qquad   \qquad \qquad  \,\,\,\,\,  \eta_0(x)  =    \csch^2 \left( y \right)  \,,  \\ \nonumber          
 \omega^2_1 &=   \frac{3}{4} m^2 \,  ,  \qquad \qquad  \,\,\,\,\,\,\,\, \eta_1(x)  =  \coth \left(y \right)
 \,  \csch \left( y\right), \\   
  \omega^2_k &=   k^2 +  m^2 \,  , \qquad  \qquad \eta_k (x)  =  e^{i k x} \, H( \coth y)/H(\sgn(x) \, C_v )  .
\end{align}
where $C_v =   \coth \left(\frac{1}{2} \log \frac{v_2}{v_1} \right) = (v_1 +  v_2)/(v_2 -  v_1)$ and  the function $H(y)$ is defined in Eq. (\ref{eq:defHy}). The 
spectrum coincides with that calculated in the previous subsection,  but the eigenfunctions are now given in terms of  $\coth y$  instead of $\tanh x$ functions.
All  the eigenfunctions  in Eq. (\ref{eq:specB}) are continuous at the origin, but their first derivatives are discontinuous. In particular, as expected, the zero energy mode is given by $\eta_0(y)   \propto d \phi_B / dx (y)$, with  
$disc \eta_{x,0} = - 2 m^2 v_1 v_2 (v_1 + v_2) / (v_2 -v_1)^2$. For the two discrete modes  the values of the coefficients $D_k$  in Eq. (\ref{eq:eqschB})  are negative, thus the delta terms represent  attractive potentials that explain the existence of bound states. 

For the continuous states, we observe that the potential in  Eq. (\ref{eq:eqschB}) is again transparent, hence  $ \eta_k (x) = e^{i kx  \pm \delta_B(k)/2}$ as $x \to \pm \infty$, and  the phase shift  computed  from Eq. (\ref{eq:specB}) becomes 

 \begin{equation}
\label{eq:deltaB}
 \delta_B(k) = -2 \arctan  \left[ \frac{3 \, k  \, m }{ m^2 - k^2}  \right] + 2 \arctan  \left[ \frac{6  \, k   \, m    \, C_v}{ m^2 (3 C_v^2 -1 ) - 4 k^2}  \right].
     \end{equation}
The first part of $\delta_B(k)$ exactly coincides with the phase shift $\delta_C(k)$ obtained in the previous subsection. Hence we can separate the contribution of the continuos modes to the kink energy in Eq. (\ref{eq:Mcont})   as $M_{cont-B} = M_{cont-C} +  \Delta M_{cont-B}$, where $M_{cont-C}$ is given in Eq. (\ref{eq:McontC}) and $ \Delta M_{cont-B}$  is obtained substituting the second part of  Eq. (\ref{eq:deltaB}) into Eq. (\ref{eq:Mcont}) and integrating by parts  resulting in 

\begin{equation}
\label{eq:DcontB}
\Delta M_{cont-B} =  \frac{3 m}{2  \pi} C_v  + \frac{3 m C_v}{  \pi}  \int_0^ \Lambda
 \frac{ \omega_k  \, [m^2  (3C_v^2 -1) + 4 k^2 ] }{ \left[ m^2  (3C_v^2 -1) - 4 k^2 \right]^2+ 9 k^2  m^2  C_v^2 } \,  \, dk,
     \end{equation}
 where $ \omega_k  = \sqrt{m^2 +k^2 } $. The denominator in the preceding integral can be factorized according to  $4 (k^2 - m^2 r_1)(k^2 - m^2 r_2)$, where
 
 \begin{equation}
\label{eq:factor}
r_{1,2}  = \frac{1}{ 8} (3C_v^2 +2 ) \pm  \frac{3}{ 8}  \sqrt{4 -3 C_v^2 } .
     \end{equation}
    The integral in Eq. (\ref{eq:DcontB})  can now  be explicitly  evaluated separating  the denominator in partial fractions. After a detailed calculation  the final result  for  $\Delta M_{cont-B} $ can be worked out as 

\begin{equation}
\label{eq:MB44}
\Delta M_{cont-B} =  \frac{3 m}{2  \pi} C_v \Bigg[ 1 + F(r_1,r_2)  + \log  \frac{2 \Lambda}{  m}    \Bigg],
     \end{equation}
     where $ F(r_1,r_2)$ is defined as
  
\begin{align}
\label{eq:FF}
 F(r_1,r_2) &= \frac{1}{2( r_1 - r_2)}  \Bigg[ \left(4r_2 + 3 C_v^2-1 \right) G(r_2) -\left(4r_1 + 3 C_v^2-1   \right) G(r_1) 
 \Bigg],   \nonumber  \\   
 G(r_i) &= \sqrt{  \frac{1 +r_i }{r_i } } \, \arctanh  \sqrt{  \frac{1 +r_i }{r_i } } , \qquad i=1,2.
   \end{align}  
  Considering that $ \langle {\cal V_C}  \rangle = m v_1 /   \lambda (v_2 - v_1)$,  the mass counter-term is worked out as

 \begin{equation}
\label{eq:MrenB}
-  \,   \frac{1}{2}  \,\delta m^2  \,  \langle { \cal V_B} \rangle  \, =   -\frac{3   \, m }{   \pi}  \, \frac{v_1 }{ v_2 - v_1} \,   \log  \frac{2 \Lambda}{  m} .
\end{equation}
  Adding the contributions of  the continuous modes in  Eqs. (\ref{eq:McontC},\ref{eq:MB44}) to the previous result  corroborates the exact cancelation of the divergent terms.  Collecting all the finite terms yields the quantum mass for the $ \phi_B$ kink

\begin{equation}
\label{eq:mBren}
    {\tilde M}_B=  \frac{m^3}{ 6 \, \lambda}\frac{Q_B^2 (3- \abs{Q_B})}{ (1- \abs{Q_B})^3  } + m \, \left( \frac{1}{ 4 \sqrt{3}  }   \, 
 +   \frac{3  }{  \pi}  \frac{v_1 }{ v_2 -v_1 }  +  \frac{3 }{2  \pi}   \, C_v  \,  F(r_1,r_2) \right) + \,  { \cal O }  (\lambda).
           \end{equation}
This  proves that the current formalism allows us to adequately analyse the discontinuities that appear in the case of kink B, giving rise to a finite value for the quantum mass of the kink.

Unlike the results obtained in the previous subsection, the stability equation and the quantum mass corrections of the $ \phi_B$ kink  Eqs. (\ref{eq:eqschB},\ref{eq:specB},\ref{eq:mBren})
 have not been previously obtained. As mentioned, the use of the reconstruction method applied to the $ \phi_C$  kink leads to either  the  $ \phi^4$   or the  $ \phi^4_G$ models. However, we propose that a simultaneous application of the reconstruction formalism to the $ \phi_C$ and $ \phi_B$ solutions will probably lead to a unique $ \phi^4_G$ field scalar theory, which is a topic that  deserves further  investigation.

%%%%%%%%%%%%%%

\section{Final considerations}\label{conclu}

This paper  analyses the properties of the $\phi^4_G$ model. In spite of the non-analytical behavior of the potential,  we prove that a systematic treatment  of the discontinuities induced in the field configuration enables  obtaining results with  clear  physical content. Kink solutions with well defined properties are obtained, and their mutual interactions as well as their quantum mass corrections are explicitly calculated. These results are of value for the study of scalar field theories in the presence of abrupt interfaces as well as for  the study  of non-equilibrium stationary steady-states \cite{Aron2020}.

At the transition point $\delta=0 $, where all the potential minima are degenerate, three kinks 
$\phi_A$, $\phi_B$, $\phi_C$ were found. The solutions  $\phi_A(x)$ and  $\phi_C(x)$ Eq. (\ref{eq:phiA}),  are equivalent to those obtained for the kink in the $\phi^4$ model, but for $\phi_B$  it is essential  to consider the effects of the potential  discontinuities  to obtain  a novel solution  and compute its  quantum kink mass  corrections (Eqs. \ref{eq:phiB} and \ref{eq:mBren}).

 The  existence of multi-soliton solutions  \cite{Gardner1967,Hirota1971,Ablowitz1973} is of great interest to the understanding of the emergence of coherent structures \cite{Scott2003,Ahlqvist2015}. Diverse  mathematical techniques have been developed to obtain such solutions \cite{Arshad2017A,Arshad2017,Hossen2018,Ullah2020}. It should be noted that in general these solutions are time dependent. In this paper it  is proven that  a stationary multi-kink  occurs near  the transition  from the degenerate  to  the  non-degenerate  potential phase.  In this region, each of the   $\phi_A(x)$, $\phi_B(x)$ and $\phi_C(x)$  kinks will be separately unstable, but they merge in a single stable multi-kink configuration because   the mutual interactions  produces a confining potential. This result was obtained in the context of  the model $\phi^4_G$,  but its  appearance can be expected  in general near a  phase transition between  a degenerate to  a non-degenerate minima potential as long as in the transition point the potential possesses at least three absolute degenerate minima \cite{Demirkaya2017}. 

 The analysis of the stability condition and the quantum corrections to the kink masses produced a modified 
Schr\"{o}dinger equation, with a reflectionless potential that also  includes  an extra delta potential term. The study of the SUSY-QM structure  \cite{Boya1989,Cooper1995} of this  equation, in which the superpotential $W(y) =  m \coth y $ is defined in terms of a piecewise variable, clearly deserves further study. Similarly, the relation of the $\phi^4$ and $\phi^4_G$ through its  connection to  the reconstruction method \cite{Bazeia2017}  should also be subject  to  further investigation.

%%%%%%%%%%%%%%

\appendix

\section{Spontaneous symmetry breaking in the generalized  $\phi^4_G$ theory}\label{Ap1}

The generalized $\phi^4_G$ potential $U(\phi)$ in  Eq. (\ref{eq:Potencial})  has two different minima: $V_1$ and  $V_2 $, and each one is degenerate due to the $Z_2$ discrete symmetry of the theory. As a consequence, the spontaneous symmetry breaking of the theory can be induced by expanding the field $\phi$ around any of the non-equivalent minima $V_i \,, \, i=1,2$;  yielding two independent Lagrangians $\mathcal{L}_i$.
Substituting the  expansion of the scalar field  as $\phi(x)=V_i+\sigma_i(x)$ into 
Eq. (\ref{eq:Lagrangian}) gives  $\mathcal{L}_i$ as 
   
\begin{equation}
\label{eq:Li}
	\mathcal{L}_i=\frac{1}{2}(\partial_{\mu}\sigma_i)(\partial^{\mu}\sigma_i)-\left(U(V_i)+\frac{m^{2}_{i}}{2}\sigma_{i}^2+\lambda_i\sigma^3_i+\lambda\sigma_i^4\right), \qquad \qquad i=1,2,
\end{equation}
where  the masses  $m_{i}$ are given in  Eq. (\ref{eq:m12}), whereas the vacuum energies  $U(V_i)$ and  the  three-point vertex  couplings $\lambda^{(3)}_i$ are worked out  as follows

\begin{align}
\label{eq:ui}
U(V_1)&=\lambda\Bigg[-\frac{\delta^{2}}{4} + \frac{5 \, \delta \, \Delta_v^{2}}{8}- \frac{\delta \, \Delta_v R}{4} + \frac{\Delta_v^{4}}{32} - \frac{\Delta_v^{3} \, R}{32} \Bigg], \nonumber \\
U(V_2)&=0, \nonumber \\
\lambda^{(3)}_1 &=\lambda\left(\Delta_v + R\right), \nonumber \\
\lambda^{(3)}_2&=2\lambda\Delta_v    ,
\end{align}
where $\Delta_v=v_2-v_1$ and  $R= \sqrt{(v_2-v_1 )^2-8\delta }$. From these expressions, the standard procedure can be followed to   extract the Feynman rules that allow  performing perturbative calculations. Clearly the results will depend on the minimum around which the perturbation is considered.  However, in the degenerate potential minima limit ($\delta=0$)  all the parameters of the  Lagrangians $\mathcal{L}_1$  and $\mathcal{L}_2$ coincide, so a perturbative calculation result  is indistinct of the selected minima. For example the evaluation of the tadpole diagram  leads to the result $ \delta m^2$ in Eqs. (\ref{eq:tadp}), already known in  literature  \cite{Dashen1974}.

%%%%%%%%%%%%%%%%%%%%%%%%%%%%%%%%%%%%%%
\section{ Solution of  the stability equation  for $\phi_B$ kink }\label{Ap2}

In this appendix we present  the   solution of the stability Schr\"{o}dinger equation Eq. (\ref{eq:eqschB})  for the fluctuations around the  $\phi_B$ kink. This equation includes the potential $ {\cal V}_B (z)  =  U_0^{ \prime  \prime}   ( \phi_B(z) ) -m^2  \, =    \, 3  \, m^2  \csch^2 \left( y \right)$ and also a delta potential contribution that appears from the fact that  $U( \phi)$  in non-analytical at  $\phi =0$. 

The potential $ {\cal V}_B (z)$   apparently presents two problems. The first one is that $ \csch^2  y $ is divergent at $y=0$, which is  why this potential, known  as the Eckart potential, has previously been used  only in three dimensional problems,  including the effect of a centrifugal barrier  \cite{Cooper1995}. However, according to the definition in  Eq. (\ref{eq:defy}) $\abs{y} \ge \log(v_2 /v_1) $, thus the potential $ {\cal V}_B (x) $  is finite for  all  the $x$ coordinate values. The other possible problem is that $ {\cal V}_B (x) $  represents a repulsive potential, so discrete bound states would not  be expected, which would exclude the existence of zero mode. Notwithstanding,  it is important to  consider the delta contribution that adds an attractive interaction producing two bound states, one of which is precisely  the zero mode. 

Introducing  the auxiliary variable  $\xi = \coth y$, Eq. (\ref{eq:eqschB})  becomes

\begin{equation}
\label{eq:eqschB2}
        \left(  \frac{d^2 }{d\xi^2} -  \frac{2 \, \xi }{(1 - \xi^2 )} \frac{d }{d\xi} +   \frac{ 2  \, \epsilon + 6(1 - \xi^2 )}{(1 - \xi^2 )^2}  + 2 \frac{ D_k}{(1 - \xi^2 )^2} \,   \delta(x)  \right) \eta_k (\xi) =  0  ,   
\end{equation}
where  $\epsilon_i = 2 (\omega_i^2 -m^2)/m^2$.
In (\ref{eq:eqschB2}) we identify the differential equation for the associated Legendre Polynomials. 
Considering $x \ne 0$ the bound-state ($ \epsilon < 0$) solutions are obtained as $P_2^{\sqrt{- 2 \epsilon}} (\xi)$. Finite  solutions at $\xi = \pm 1$ are obtained only when $ \sqrt{-2 \epsilon} = 1,2 $,  yielding two discrete  modes $\omega_0^2 = 0$ and  $\omega_1^2 = \frac{3}{4}m^2 $, with their corresponding eigenfunctions  $\eta_0(x) \propto P_2^2 (\xi)$ and  $ \eta_1(x) \propto  P_2^1 (\xi)$, that are explicitly given in Eq. (\ref{eq:specB}). We  verify  that both $\eta_0(x)  $ and $\eta_1(x)  $ are continuous  at $x = 0$, and the discontinuity of their first derivatives  at the origin  produce   factors 
$ D_ 0 = - 2 m ( v_1+ v_2)/(v_2 - v_1) $ and   $ D_ 1 = -4 m v_1  v_2/(v_2^2 - v_1^2)$ required to satisfy  the condition imposed to the solutions by the $\delta(x)$ potential term in Eqs. (\ref{eq:eqschB},\ref{eq:eqschB2}).

In the scattering regime $ \epsilon_k = 2 k/m$, and  the solution to Eq. (\ref{eq:eqschB2}) is given by 

\begin{equation}
\label{eq:eqsolk} 
      \eta_k(x) \propto      P_2^{i 2 k /m} (\xi) \propto e^{i k x} \null_2F_1 \left(-2,3,1 - 2 i  k /m, 1 - \xi\right)
      \propto e^{i k x}  \, H ( \coth y ),
\end{equation}
where  $P_2^{i 2 k /m} (\xi)$ is expressed in terms of the   hypergeometric function\cite{Morse1953}  $   \null_2F_1$   that reduces to the last term in  Eq. (\ref{eq:eqsolk}) with $H(\xi)$ defined in Eq. (\ref{eq:defHy}). Finally taking into account  that  $y $ defined in  Eq. (\ref{eq:defHy}) is piecewise, the normalization constants for $  \eta_k(x)$ have to be  separately selected  for $x < 0$ and  $x > 0$ in order to enforce the continuity condition at the origin, giving rise to the following result for the continuous eigenfunctions 

\begin{equation}
\label{eq:eqsolk2} 
      \eta_k (x)  =  e^{i k x} \,  \frac{H( \coth y)}{H(\sgn(x) \, C_v )}   .
     \end{equation}

 %%%%%%%%%%%%%%%%%%%%%%%%%%%%%%%%

%%%%%%%%%%%%%%%%%%%%%   ------- %%%%%%%%%%%%%%%%%%%%%
%                   REFERENCIAS                     %

%\bibliographystyle{abbrv}	
\bibliography{References.bib} 
%%%%%%%%%%%%%%%%%%%%%   ------- %%%%%%%%%%%%%%%%%%%%%

%%%%%%%%%%%%%%%%%%%%%%%%%%%%%%%%
 
 %%%%%%%%%%%%%%%%%%%%%%%%%%%%%%%%

 %%%%%%%%%%%%%%%%%%%%%%%%%%%%%%%%

%%%%%%%%%%%%%%%%%%%%%%%%%%%%%%%%
 %%%%%%%%%%%%%%%%%%%%%%%%%%%%%%%%

\end{document}